\documentclass[aps,prc,twocolumn,nofootinbib,showpacs]{revtex4}

\usepackage{graphicx}
\usepackage{dcolumn}
\usepackage{bm}       
\usepackage{latexsym} 
\usepackage{longtable}
\usepackage{array}
\usepackage{rotating}
\begin{document}
\title{$B(E2)$ strength ratio of one-phonon $2^+$ states of $^{94}$Zr from electron scattering at low momentum transfer}

\author{A.~Scheikh Obeid}
\author{S.~Aslanidou}
\author{J.~Birkhan}
\author{A.~Krugmann}
\author{P.~von~Neumann-Cosel}\email{vnc@ikp.tu-darmstadt.de}
\author{N.~Pietralla}
\author{I.~Poltoratska}
\author{V.~Yu.~Ponomarev}

\affiliation{Institut f\"ur Kernphysik, Technische Universit\"at Darmstadt, 
                64289 Darmstadt, Germany}

\date{\today}

\begin{abstract}
\noindent
{\bf Background:} 
The $B(E2)$ transition strength to the $2^+_2$ state in $^{94}$Zr was initially reported to be larger by a factor of 1.63 than the one to the $2^+_1$ state from lifetime measurements with the Doppler-shift attenuation method (DSAM) using the $(n,n'\gamma)$ reaction [E. Elhami {\it et al.}, Phys. Rev. C \textbf{75}, 011301(R) (2007)].
This surprising behavior was recently revised in a new measurement by the same group using the same experimental technique leading to a ratio below unity as expected in vibrational nuclei. \\ 
{\bf Purpose:} The goal is an independent determination of the ratio of $B(E2)$ strengths for the transitions to the $2^+_{1,2}$ states of $^{94}$Zr with inelastic electron scattering. \\
{\bf Method:} The relative population of the $2^+_{1,2}$ states in the ($e,e'$) reaction was measured  at the S-DALINAC in a momentum transfer range $q = 0.17 - 0.51$ fm$^{-1}$ and analyzed in plane-wave Born approximation with the method described in A. Scheikh Obeid {\em et al.}, Phys. Rev. C \textbf{87}, 014337 (2013). \\   
{\bf Results:} The extracted $B(E2)$ strength ratio of $0.789(43)$ between the excitation of the $2^+_1$ and $2^+_2$ states of $^{94}$Zr is consistent with but more precise than the latest $(n,n'\gamma$) experiment. 
Using the $B(E2)$ transition strength to the first excited state from the literature a value of 3.9(9) W.u.\  is deduced for the $B(E2;2^+_2\rightarrow0^+_1$) transition. \\
{\bf Conclusions:} The electron scattering result independently confirms the latest interpretation of the different $(n,n'\gamma$) results for the transition to $2^+_2$ state in $^{94}$Zr. 
\end{abstract}

\pacs{21.10.Re, 23.20.Js, 25.30.Dh, 27.60.+j}
                           
\maketitle

The investigation of collective valence shell excitations  provides a direct way to address the interaction between valence protons and neutrons in many-body fermionic quantum system like atomic nuclei~\cite{pie08}. 
In even-even vibrational nuclei the interaction is such that the basic proton and neutron quadrupole transitions form collective isoscalar and isovector excitations \cite{wal11}. 
The isovector excitation is the so-called quadrupole mixed-symmetry state~\cite{iac87}. 
The isoscalar vibration is generally lowest in energy and expected to show the largest ground state (g.s.) transition strength of all quadrupole excitations in the valence shell, {\it i.e.}\
$B(E2;2^{+}_{1} \rightarrow 0^{+}_{1}) > B(E2;2^{+}_{i} \rightarrow 0^{+}_{1})$, $i>1$.
In $^{94}$Zr, the $2^+_2$ state was identified as the mixed-symmetry state based on its decay behavior \cite{elh07,elh08} and the magnetic moment \cite{wer08}. 
Elhami {\it et al.}~\cite{elh07} also reported an inversion of the ground state (g.s)  transition strengths, {\it i.e.},  the  $B(E2;2^{+}_2 \rightarrow 0^+_1)$ transition strength was claimed to be significantly larger than that of the $2^+_1$ state. 
This unexpected behavior was confirmed in shell-model calculations \cite{sie09}.
However, it was pointed out that the discrepancy was theoretically less pronounced and sensitively depends on the amplitude ratio of the main proton and neutron configurations forming the mixed-symmetry state.
  
Recently, a significantly smaller result for the $B(E2;2^{+}_2 \rightarrow 0^+_1)$ transition strength has been reported by the same group using the same experimental technique, viz. lifetime measurements with DSAM after population with the $(n,n'\gamma)$ reaction  
\cite{cha13}.
The difference in the results has been related to the chemical properties of the scattering targets used in those experiments. 
It has been shown that the measured lifetimes differ for amorphous (Zr(OH)$_{4}$) and crystalline (ZrO$_{2}$) material and the effect depends on the particle size \cite{pet13}.
Shorter lifetimes have been observed for samples composed of smaller particles.  
The new value $B(E2;2^{+}_2 \rightarrow 0^+_1) = 3.9(3)$ W.u. is now below the corresponding value of 4.9(11) W.u. for decay from the $2^+_1$ state.

Because of the general implications of Ref.~\cite{pet13} on DSAM measurements an independent check of this result is desirable.
Here, we report on a new measurement of the ratio of the $B(E2)$ strengths from electron scattering.
It has recently been shown in the neighboring isotope $^{92}$Zr \cite{sch13} that the ratio of $B(E2)$ strengths of one-phonon transitions can be directly extracted from a relative analysis in plane-wave Born approximation (PWBA).
While a PWBA analysis is generally not sufficient to describe the electron distortion in heavy nuclei, the Coulomb effects cancel in the relative analysis.
Thus, the ratio can be determined with high precision because it is largely independent of many contributions to the systematic error.
 
\begin{figure}
\begin{center}
\includegraphics[width=8.5cm]{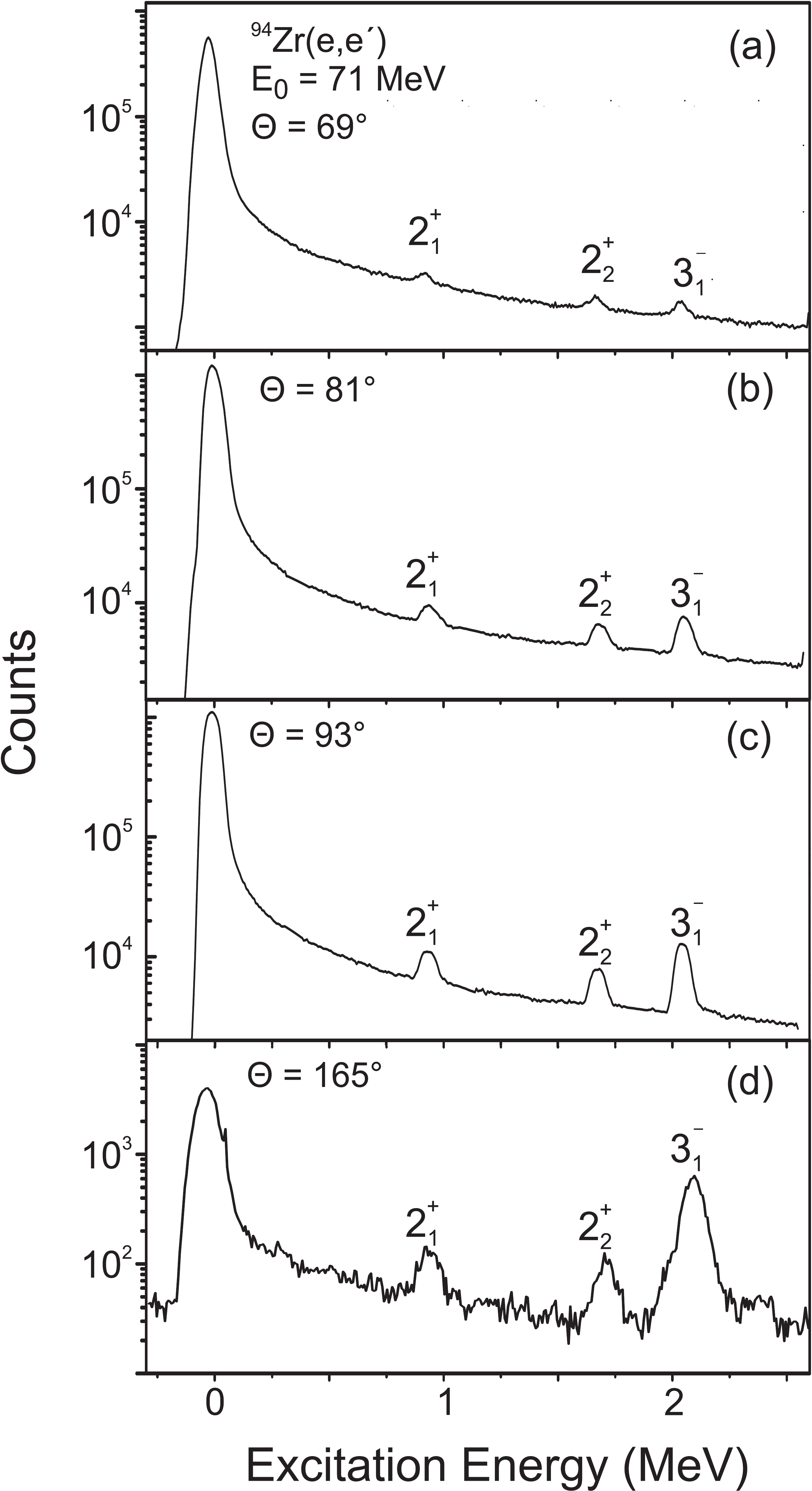}
\caption{Four electron scattering spectra of the $^{94}$Zr(e,e$'$) reaction at incident electron energy E$_{0}$ = 71~MeV and electron scattering angles $\Theta$ = $69^{\circ}$, $81^{\circ}$, $93^{\circ}$ and $165^{\circ}$.}
\label{Zr94_4spec}
\end{center}
\end{figure}

The experiment has been carried out at the Darmstadt superconducting electron linear accelerator S-DALINAC. 
The high-resolution spectrometer Lintott with its focal-plane detector system based on four single-sided silicon strip detectors, each providing 96 strips with a thickness of 500~$\mu$m and a pitch of 650~$\mu$m \cite{len06}, was used. 
Measurements were performed for $^{94}$Zr with an incident electron beam energy $E_{0} = 71$ MeV and beam currents ranging from $0.5$ to $2$ $\mu$A. The $^{94}$Zr target had an isotopic enrichment of $96.07\%$ and a thickness of 10 mg/cm$^{2}$. 
Data were taken at four different scattering angles $\theta$ = $69^{\circ}$, $81^{\circ}$, $93^{\circ}$ and $165^{\circ}$ covering the maximum of the E2 form factor. 
The average energy resolution was about 60 keV (full width at half maximum). 
The electron-scattering spectra are shown in Fig.~\ref{Zr94_4spec}.
The prominent peaks correspond to the elastic line, the collective one-phonon $2^{+}_{1,2}$ and $3^{-}_{1}$ states. 
The spectra were energy calibrated with the excitation energies of these states taken from Ref.~\cite{elh08}.

Peak areas $A$ of the transitions were obtained from a spectrum decomposition using the line shape described in Ref.~\cite{hof02}. 
The peak area ratios for the $2^+_{1}$ and $2^+_{2}$ states are given in Table~\ref{tab:ff} as a function of the squared momentum transfer 
 \begin{equation}
\label{eqn:q}
q = \frac{1}{\hbar c}\sqrt{2E_{\rm 0}\left(E_{\rm 0}-E_{\rm x}\right)\left(1-cos~\theta\right)+E^{\rm 2}_{\rm x}}.
\end{equation}
\begin{ruledtabular}
\begin{table}[t]
\caption{
Peak area ratios of electroexcitation of the $2^+_{1,2}$ states in $^{94}$Zr at $E_0 = 71$ MeV for different momentum transfers and the corresponding kinematical correction factor $R_F$ defined in Ref.~\cite{sch13}.
}
\label{tab:ff}
\begin{tabular}{c c c c}
$\Theta$&$q^{2}$ (fm$^{-2}$) &$2^{+}_{2}$/$2^{+}_{1}$& $R_F$ \\
\hline
69$^{\circ}$&$0.17$  & $0.929(103)$    &  1.011 \\
81$^{\circ}$&$0.22$  & $0.801(34)$    &  1.011 \\
93$^{\circ}$&$0.27$  & $0.784(25)$     &  1.011 \\
165$^{\circ}$&$0.51$ & $0.921(95)$  &  1.006 \\
\end{tabular}
\end{table}
\end{ruledtabular}
It was shown in Ref.~\cite{sch13} that in a PWBA analysis the ratio of the peak areas of the $2^+_{1}$ and $2^+_{2}$  states can be expressed as 
\begin{eqnarray}
\label{eqn:ratioDR}
& R_{\rm F}(q)\sqrt{\frac{\displaystyle A_{2}}{\displaystyle A_{1}}} \approx  \sqrt{\frac{\displaystyle B(E2,k_2)}{\displaystyle B(E2,k_1)}}  \nonumber \\ & \times \left(\frac{1- \frac{\displaystyle q_2^2}{\displaystyle 14} \displaystyle \left(R_{{\rm tr},1} + \Delta R \right)^2 + \frac{\displaystyle q_2^4}{\displaystyle 504} \displaystyle \left(R_{{\rm tr},1} + \Delta R \right)^4}{1- \frac{\displaystyle q_1^2}{\displaystyle 14}  \left(R_{{\rm tr},1}\right)^2 + \frac{\displaystyle q_1^4}{\displaystyle 504}  \left(R_{{\rm tr},1}\right)^4} \right),
\end{eqnarray}
%
%
where the indices $1,2$ indicate the transitions to the $2^+_1$ and $2^+_2$  
state, respectively. 
Here, $R_{{\rm tr},i}$, $i=1,2$ denote the transition radii defined in Eq.~(5) of Ref.~\cite{sch13}, $\Delta R = R_{{\rm tr},2} - R_{{\rm tr},1}$, and $R_F$ stands for a kinematical correction factor defined in Ref.~\cite{sch13}.
At the photon point $q = k = E_{\rm x}/\hbar c$, the r.h.s.\ of Eq.~(\ref{eqn:ratioDR}) reduces to the square root of the $B(E2)$ excitations strengths to the $2^+$ states of interest.

\begin{figure}[tbh]
\begin{center}
\includegraphics[width=8.5cm]{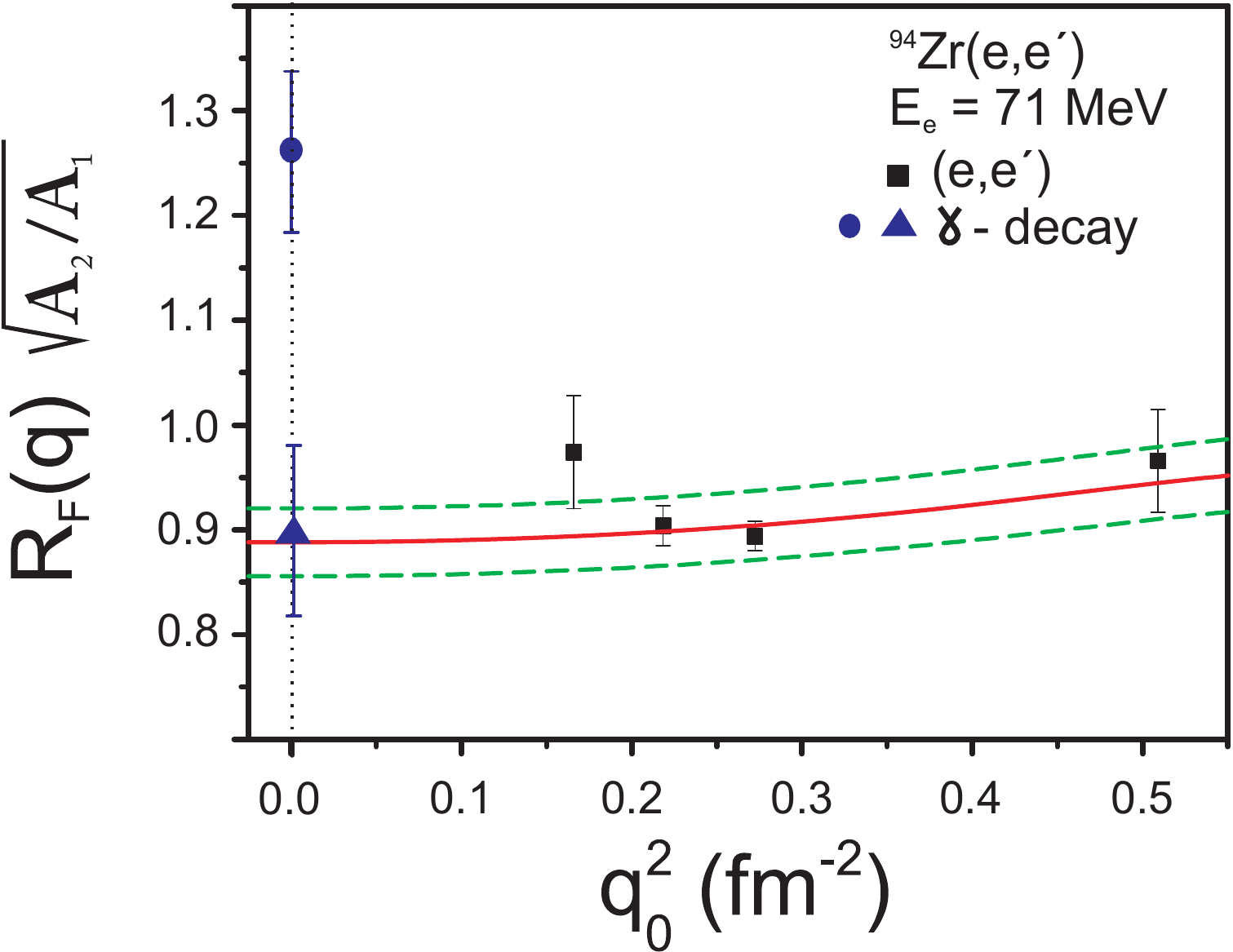}
\caption{(color online). 
Kinematically corrected square root ratio of peak areas of the $2^+$ MSS and FSS (solid squares) of $^{94}$Zr as a function of the squared elastic momentum transfer $q_{0}$. 
At the photon point indicated by the vertical dashed line, this observable equals the square root of the corresponding $B(E2)$ transition strengths. 
Additionally, data points deduced from the ratio of $B(E2)$ strengths obtained from $\gamma$-decay lifetime measurements are shown as blue full circle \cite{elh07} and full triangle \cite{pet13}, respectively. 
The red solid curve is a fit of Eq.~(\ref{eqn:ratioDR}) with $1\sigma$ error bars indicated by the green dashed curves.}
\label{94Zr-Fit}
\end{center}
\end{figure}

Figure~\ref{94Zr-Fit} shows a plot of the $R_{\rm F}\sqrt{A_2/A_1}$ values as a function of the squared elastic momentum transfer 
$$q_0 = \frac{1}{\hbar c} \sqrt{2E_0^2(1-\cos\theta)}.$$
In the fit of Eq.~(\ref{eqn:ratioDR}) to the data we have fixed $R_{{\rm tr},1}$ to the same value as for $^{92}$Zr, viz.\ $R_{{\rm tr},1}=5.6$~fm.
It was shown in Ref.~\cite{sch13} that the fit results are independent of particular choice of $R_{{\rm tr},1}$ over a wide parameter range (at least  $\pm 1$ fm$^{-1}$).  
In this way the number of parameters is reduced to two, viz.\ the ratio of $B(E2)$ strengths and $\Delta R$. 

A $\chi^{2}$-minimization of Eq.(\ref{eqn:ratioDR}) to the data then yields ${B(E2;2^{+}_{2})}/{B(E2;2^{+}_{1})}= 0.789(43)$ in contradiction to the result  $1.63(37)$ from Ref.~\cite{elh07} but in good agreement with Ref.~\cite{pet13}, who find $0.79(10)$ (see also Fig.~\ref{94Zr-Fit}). 
Taking the value $B(E2;2^+_1)=4.9(11)$ W.u. from Ref.~\cite{ram01}, we extract $B(E2;2^{+}_2) = 3.9(9)$~W.u. from the measured ratio.
%
The difference of the transition radii of the two states entering as second fit parameter in Eq.~(\ref{eqn:ratioDR}) is consistent with zero ($\Delta R$ = -0.28(42) fm).
The implications of this result for a possible mixed-symmetry character of the $2^+_2$ state in $^{94}$Zr will be discussed elsewhere \cite{wal13}.

To summarize, we report on inelastic electron scattering measurements of the $2^+_{1,2}$ states of $^{94}$Zr at the \mbox{S-DALINAC}.
In a PWBA analysis \cite{sch13}, where a direct relation beween the form factors and the transition strengths can be established, the ratio of the $B(E2)$ strengths is determined with high precision.
The value is consistent with Ref.~\cite{pet13} and independently confirms the result of their reanalysis of the lifetime of the $2^+_2$ state.
The small uncertainty of the PWBA analysis (because of the cancellation of systematic errors in a relative measurement) can be used in future to improve the uncertainties of the absolute $B(E2)$ values. 
 
We thank R.~Eichhorn and the S-DALINAC crew for their commitment in delivering electron beams.
This work has been supported by the DFG under grant No.\ SFB 634.

\end{document}